\documentclass[preprint,showpacs,preprintnumbers,amsmath,amssymb,aps]{revtex4}

\usepackage{graphicx}
\usepackage{dcolumn}
\usepackage{bm}

\begin{document}

\newcommand{\LSMO}{La$_{\mbox{\scriptsize
7/8}}$Sr$_{\mbox{\scriptsize 1/8}}$MnO$_{\mbox{\scriptsize 3}}$ \/}
\newcommand{\LSMOo}{La$_{\mbox{\scriptsize
7/8}}$Sr$_{\mbox{\scriptsize 1/8}}$MnO$_{\mbox{\scriptsize 3}}$}
\newcommand{\LSMOx}{La$_{\mbox{\scriptsize
1-x}}$Sr$_{\mbox{\scriptsize x}}$MnO$_{\mbox{\scriptsize 3}}$ \/}
\newcommand{\lii}{La L$_{\mbox{\scriptsize II}}$-edge \/}
\newcommand{\liio}{La L$_{\mbox{\scriptsize II}}$-edge}
\newcommand{\mnk}{Mn K-edge \/}
\newcommand{\mnko}{Mn K-edge}
\newcommand{\sigpi}{$\sigma\pi$}
\newcommand{\sigsig}{$\sigma\sigma$}
\newcommand{\tco}{T$_{\mbox{\scriptsize CO}}$}
\newcommand{\tjt}{T$_{\mbox{\scriptsize JT}}$}
\newcommand{\tc}{T$_{\mbox{\scriptsize C}}$}
\newcommand{\tilt}{I$_{\mbox{\scriptsize 300}}$}
\newcommand{\gamm}{$1/\gamma^2$}
\newcommand{\LPSMO}{(La$_{\mbox{\scriptsize 1-y}}$Pr$_{\mbox{\scriptsize y}}$)$_{\mbox{\scriptsize 7/8}}$Sr$_{\mbox{\scriptsize 1/8}}$MnO$_{\mbox{\scriptsize 3}}$}

%


\title{Octahedral tilts and electronic correlations in \LSMO}

\author{J. Geck$^{1,2}$}
\author{P. Wochner$^2$}
\author{S. Kiele$^{1,3}$}
\author{P. Reutler$^1$}
\author{A. Revcolevschi$^4$} 
\author{B. B\"uchner$^1$}
%
%

\affiliation{$^1$Leibniz Institute for Solid State and Materials Reasearch  IFW Dresden, Helmholtzstr. 20, 01069 Dresden, Germany}

\affiliation{$^2$Max-Planck-Institut f\"ur Metallforschung, Heisenberg Str. 3, 70569 Stuttgart, Germany}

\affiliation{$^3$Hamburger Synchrotronstrahlungslabor HASYLAB at
 Deutsches Elektronen-Synchrotron DESY, Notkestr. 85, 22603 Hamburg,
 Germany}

\affiliation{$^4$Laboratoire de Physico-Chimie de l'Etat Solide, Universit\'e
  de Paris-Sud, 91405 Orsay Cedex, France}
\date{Received: \today}
%
\begin{abstract}

We present a resonant x-ray scattering study of the octahedral tilt order 
between 50\,K and 310\,K in La$_{7/8}$Sr$_{1/8}$MnO$_3$. At the \lii the resonant (300) reflection probes cooperative tilts of the
MnO$_6$-octahedra in this material, as verified by a model caclulation as well as a LDA+U study. The investigation of the octahedral
tilts as a function of temperature and the comparison to the lattice parameters, the magnetization and the superlattice reflections
related to charge and/or orbital order reveal an intimate coupling between electronic and tilt degrees of freedom  in \LSMOo.

\end{abstract}

\maketitle

\subsection{Introduction}\label{intro}

The investigation of the electronic correlations in doped transition metal oxides is an active field of condensed matter research. In
particular, the interplay between structural distortions and charge ordering phenomena has attracted a lot of attention. A prominent
example for the coupling between structure and charges is the influence of the octahedral tilts  on the stripe order in doped cuprates;
i.e. the pinning of charge stripes by a potential due to tilts of the CuO$_6$-octahedra \cite{TranquadaNature95,BuechnerPRL94}. The
static stripe order which develops in doped nickelates
\cite{TranquadaPRL94}  
might also be stabilized by a similar mechanism, namely by the cooperative tilting of the NiO$_6$-octahedra. Recently, it has been
recognized that octahedral tilt order also plays an important role for the physics of doped manganites. In particular, there are
indications that the tilting of the MnO$_6$-octahedra in LaMnO$_3$ couples to the orbital degree of freedom
\cite{ZimmermannPRB01,MizokawaPRB99}.

In this article we present a resonant x-ray scattering study of the octahedral tilt order in \LSMO which shows several phase transitions
as a function of temperature \cite{UhlenbruckPRL99,KlingelerPRB01}: Upon cooling an orbital ordered state similar to that of LaMnO$_3$
develops at \tjt$\simeq 280$\,K, where an antiferrodistortive ordering of  Jahn-Teller distorted MnO$_6$-octahedra occurs
\cite{KawanoPRB96,MurakamiPRL98}. Decreasing the temperature down to \tc\,$\simeq$\,180\,K leads to the onset of ferromagnetic order,
yielding a change of the electrical resistivity from insulating to metal-like in agreement with the double exchange picture. Further
cooling leads to a metal-insulator transition at \tco$\simeq$155\,K, where a charge and orbital ordered ferromagnetic insulating (FMI)
phase is established at low temperatures \cite{EndohPRL99,YamadaPRB00,KorotinPRB00,MizokawaRPRB00,GeckNJP04}. This FMI phase contradicts
a bare double exchange model, and there is strong evidence that pronounced correlations between magnetism, charges, structure and orbital
degrees of freedom stabilize an orbital polaron lattice below \tco\/ \cite{GeckPRL05}.

These results together with the aforementioned connection between the orbital order and the octahedral tilts found in LaMnO$_3$, raise
the question whether octahedral tilts are important  for the stabilization of the various phases in \LSMO as well. In order to elucidate
the role of the tilt ordering in \LSMO we have performed Resonant X-ray Scattering (RXS) at the \liio\/ which has been shown to be highly
sensitive to changes in the octahedral tilt order \cite{ZimmermannPRB01,BenedettiPRB01}.
\begin{figure}[t!]
\center{
\resizebox{0.7\columnwidth}{!}{%
   \includegraphics[clip, angle=-0]{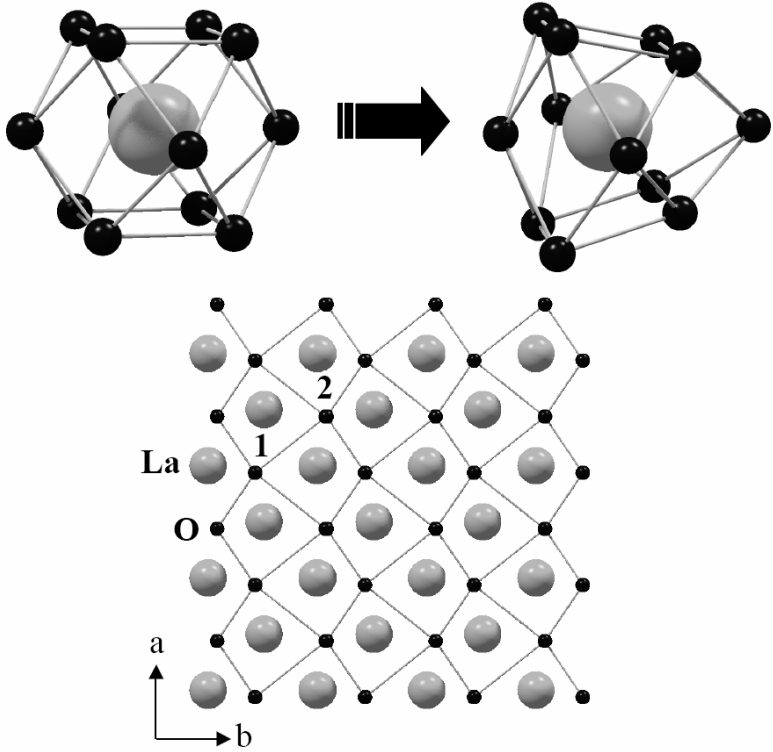}
}}
\caption{{\it Top:} Deformation of the oxygen cage around a lanthanum
   ion in the orbital ordered phase of LaMnO$_3$ (the edges of the
cage do not represent
chemical bonds).
{\it Bottom :} The tilt order in the orbital ordered phase creates
lanthanum sites with differently  distorted local environments in the
orthorhombic
{\it ab}-plane (sublattices 1 and 2).}
\label{fig:1}
\end{figure}

\subsection{Experimental technique}

By tuning the x-ray energy to the \lii the resonant scattering becomes sensitive to changes of the  octahedral tilt order. This
sensitivity stems from the fact that at the \lii the resonant scattering process involves  $2p \rightarrow 5d$ transitions in the La
sub-bands; i.e. the creation of virtual photoelectrons in an intermediate $5d$-state. As the charge distribution of the $5d$-states is
extended, the influence of the crystal field is large. Therefore, the resonant scattering strongly depends on the local environment of
the resonant scatterer. As demonstrated in Fig.\,\ref{fig:1}, the GdFeO$_3$ type tilt order in the cooperative Jahn-Teller distorted
phase of LaMnO$_3$ leads to a strong deformation of the oxygen cage around the La-site. Moreover, the tilt order of the MnO$_6$-octahedra
creates sublattices  of La-sites with differently distorted local environments (1 and 2 in Fig.\,\ref{fig:1}). This makes it possible to
observe symmetry forbidden reflections at the \liio, which reflect changes of the octahedral tilt order, as described in more detail in
the following Sec.\,\ref{pheno}.
%

The RXS experiment has been performed using a vertical scattering geometry at the wiggler beamline W1 at HASYLAB. This beamline  is
equipped with a Si(111) double monochromator which provides an energy resolution of about 2\,eV at the \liio. The sample was mounted on
the cold finger of a closed cycle cryostat which was itself mounted on a standard Eulerian cradle. The incident slits have been chosen in
order to give a polarization \mbox{$P=I^{\sigma}/(I^{\sigma}+I^{\pi})=0.93$}\/ of the incident beam, where $\sigma$ and $\pi$ refer to
polarization directions perpendicular and parallel to the scattering plane, respectively. For the polarization analysis at the \lii \/
the (004) reflection of  graphite with a scattering angle of $2\Theta\simeq77.61^{\circ}$ has been used, leading to a cross talk between
\sigsig\/ and \sigpi\/ channels of 6\,\% as determined from the (400) Bragg reflection. The \LSMO \/ single crystals have been grown
using the traveling floating zone method \cite{ReutlerCG02} and for the RXS experiment a sample with a polished (100)-surface has been
prepared (surface roughness $\simeq5$\,$\mu\mbox{m}$).

\subsection{Model calculation}\label{pheno}

The observation of symmetry forbidden reflections in LaMnO$_3$ for photon energies around the \liio, can be understood in terms of a
model calculation based on a purely ionic picture: When the x-ray energy is tuned to the \lii\/ the scattering process involves
$2p\rightarrow 5d$ transitions. Adopting the dipol-dipol approximation the relevant term of the atomic form factor $f$ near the \lii can
be described by a second rank tensor $\hat f$ with respect to the beam polarization, which is given by
\cite{IshiharaPRB00,TakahashiJPSJ99}:
\begin{displaymath}
\hat f^{(\alpha,\beta)}=\frac{1}{m} \sum_{j} \frac{
\langle 2p|p_{\alpha}|5d_j\rangle\langle5d_j|p_{\beta}|2p\rangle}
{E(5d_j)-E(2p)-\hbar \omega - i \Gamma/2}
\end{displaymath}
Here, $|2p\rangle$ and $|5d_j\rangle$ denote the initial $2p$ state and the $5d$ intermediate states, respectively. $E(2p)$ and $E(5d_j)$
are the corresponding energies, $m$ is the (reduced) electron mass, $p_{\alpha,\beta}$ ($\alpha,\beta=1,2,3$) represents the components
of the momentum operator and $\Gamma\sim 1$\,eV is the life time broadening width of the $2p$ core hole.
Since the asymmetric unit of the LaMnO$_3$ structure  contains only a single lanthanum site, the structure factor tensor used for the
kinematic calculation of the diffracted intensities reads (temperature effects are neglected)~\cite{KirfelRXSBook}
\begin{displaymath}
\hat F({\bf Q})=\sum_g \hat f_g e^{i\,{\bf Q}\cdot {\bf r}_g}\quad.
\end{displaymath}
In the above equation the sum runs over the symmetry operations $g$ of the space group, $\hat f_g=R_g\, \hat f \, R_g^t$ is the scattering
factor tensor transformed by the rotational part $R_g$ of $g$, ${\bf r}_g$ is the position vector
of the corresponding symmetry related lattice site 
and {\bf Q} stands for the scattering vector. Note, that due to the above transformation of $\hat f$ the form factor tensor of the
lattice sites 1 and 2 in Fig.\,\ref{fig:1} is different. For a {\it Pbnm}-symmetry, which we refer to throughout this paper,  the
intensities of the 'forbidden' (300) reflection in the \sigpi- and the \sigsig-channel are \cite{MorgenrothRXSBook}
\begin{eqnarray*}
I_{\sigma\pi} & \propto & (f_{12} \cos \vartheta \cos \Phi)^2 \quad
{\rm and}\\
I_{\sigma\sigma} &= & 0,
\end{eqnarray*}
where $\vartheta$ is the Bragg angle and $\Phi$ is the azimuthal angle, which describes the rotation of the sample around the scattering
vector \cite{MurakamiPRL98}. In the ionic picture the matrix element $f_{12}$ can then be calculated using the Wigner-Eckhart theorem, in
order to get an idea of its dependence on octahedral tilts and distortions. For a $2p_{1/2}$ initial state and the $5d$ intermediate
states of $t_{2g}$ and $e_g$ symmetry it follows: (the same result holds for a $2p_{3/2}$ initial state)
\begin{eqnarray*}
f_{12}&=& \frac{|M|^2}{E(5d_{yz})-E(2p)-\hbar \omega - i \Gamma/2}\\
&-&\frac{|M|^2}{E(5d_{xz})-E(2p)-\hbar \omega - i \Gamma/2} \quad\\
&\propto& |M|^2 \Delta \times \mathcal{L}_{\Delta}(\hbar \omega)
\end{eqnarray*}
($|M|^2$=const. has the unit of an energy, $\Delta=E(5d_{xz})-E(5d_{yz})$, and $\mathcal{L}_{\Delta}(\hbar \omega)$ is a Lorentzian
centered at $\Delta$). Since the $t_{2g}$-symmetric $5d_{xz}$- and $5d_{yz}$-orbitals are degenerate in a cubic crystal field, the
intensity of the (300) vanishes in this case ($\Delta=0$).
But octahedral tilts and distortions lower the symmetry of the local environment, thereby lifting orbital degeneracy ($\Delta\neq0$). In
this case $I_{\sigma\pi}$  does not vanish and the (300) reflection is observed at the \liio. As a result, the ionic model yields the
occurrence of the resonant (300) reflection at the \lii for non-vanishing octahedral tilts and/or distortions with three important main
characteristics, namely (i) pure \sigpi-scattering, (ii) a $\sin^2$-azimuthal dependence and (iii) a resonant increase of the intensity
at the La-edges.

However, the ionic picture has also its limitations. More specifically, the dependence of $f_{12}$ on the values of the tilt angles and
the size of the distortions can not be calculated accurately. This is because band structure effects, which are relevant for the
spatially extended La:5d intermediate states, are not taken into account in this description. Therefore, the ionic localized valence
charge model should only be used as a guide, whereas a rigorous discussion should always refer to the full band-structure calculation.

In a previous publication, the effect of structural distortions on the (300) reflection at the La-edges has been calculated by means of
LDA+U, which provides an appropriate description for the effects of the band structure mentioned above \cite{BenedettiPRB01}. Besides the
octahedral tilts also distortions of the octahedra have been investigated. But the LDA+U calculations show that the intensity of the
(300) reflection at the \lii is mainly given by octahedral tilts, whereas the contribution due to distortions of the octahedra has been
found to be negligible.

To conclude, the above results show that the resonant scattering at the (300) position in the \sigpi-channel at the \lii probes  the tilt
structure of \LSMO.

\subsection{Results}

\begin{figure}[t!]
\center{
\resizebox{0.95\columnwidth}{!}{%
   \includegraphics[clip, angle=-90]{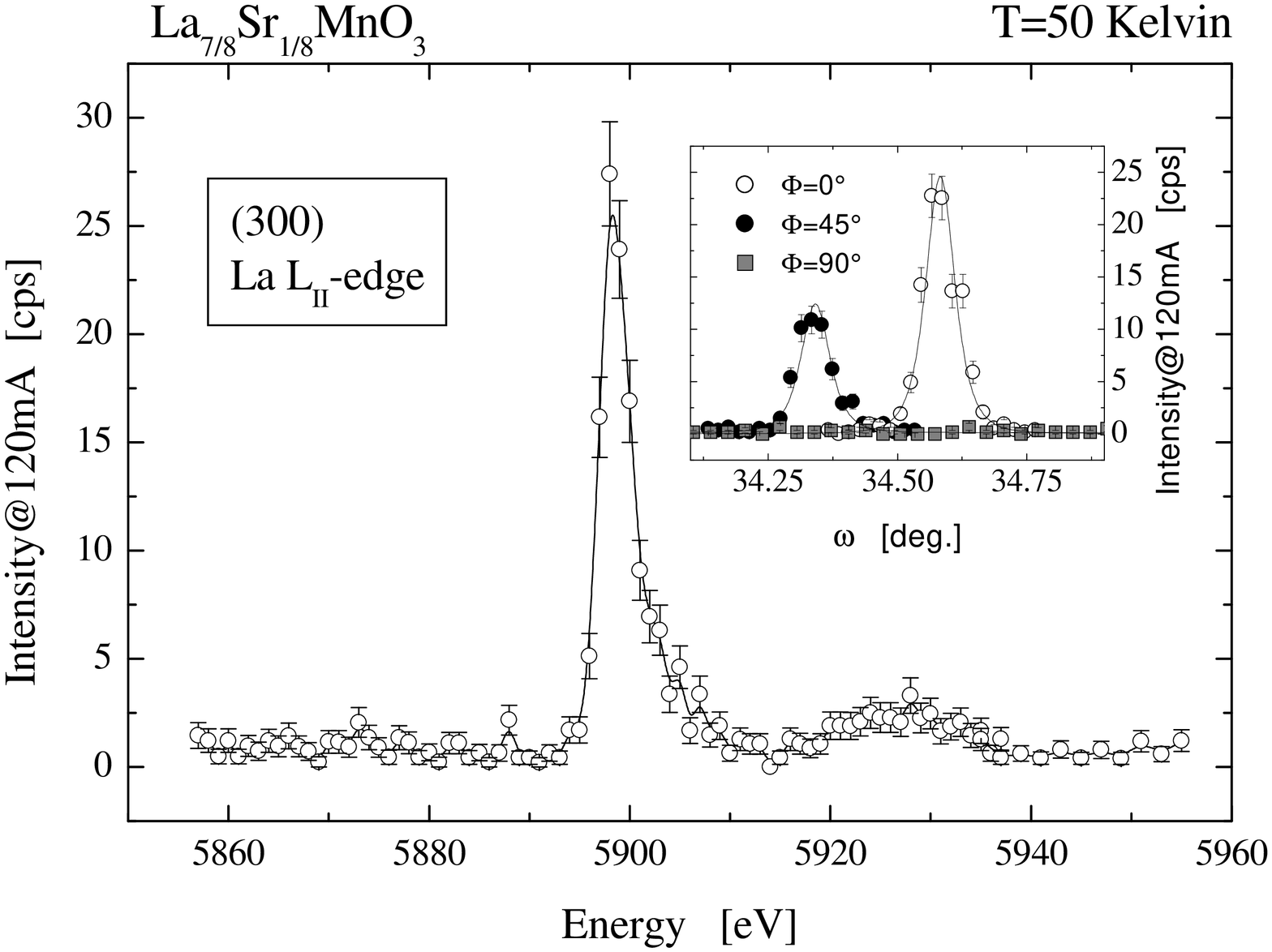}
}} \caption{ Energy dependence of the (300) reflection around the \lii in the FMI-phase  at 50K. The azimuthal dependence of a rocking
scan at the same temperature and at  5987\,eV is shown in the inset. These measurements have been performed without analyzer.}
\label{fig:2}
\end{figure}

\subsubsection{Energy and azimuthal dependence}

Figure\,\ref{fig:2} displays the energy dependence of the (300) reflection  around the \lii at 50\,K in the FMI phase. The energy
dependence displays a sharp (FWHM$\simeq $ 5\,eV) but asymmetric peak at 5897\,eV and a broad increase of the resonant intensity around
5926\,eV.  Above 5940\,eV and below 5890\,eV the intensity at the (300) position vanishes.  The observed  features of the energy
dependence shown in Fig.\,\ref{fig:2} resemble the calculated energy dependence at the \lii obtained by the LDA+U calculations mentioned
above. In the inset of Fig.\,\ref{fig:2} the pronounced azimuthal dependence of a rocking scan through the (300) reflection at 5897\,eV
can clearly be observed. Referring to the inset, the case $\Phi=0^{\circ}$  corresponds to a polarization of the incident beam parallel
to the $a,b$-direction, whereas this polarization is perpendicular to the $ab$-plane for $\Phi=90^{\circ}$. Obviously, the intensity is
continuously reduced to zero by increasing $\Phi$ from $0^{\circ}$ to $90^{\circ}$, excluding multiple scattering as the origin for the
intensity at the (300)-position. The polarization analysis performed using the (002) reflection of
graphite  
reveals that, within the experimental errors, the resonant scattering at the \lii occurs only in the \sigpi-channel. These observations
are in nice agreement with the results of the model calculation presented in Sec.\,\ref{pheno} and the predictions of the LSDA+U study
\cite{BenedettiPRB01}, verifying that the observed resonant (300) reflection in the \sigpi-channel probes the octahedral tilts in \LSMO.

\begin{figure}[t!]
\center{
\resizebox{0.8\columnwidth}{!}{%
   \includegraphics[clip,angle=-90]{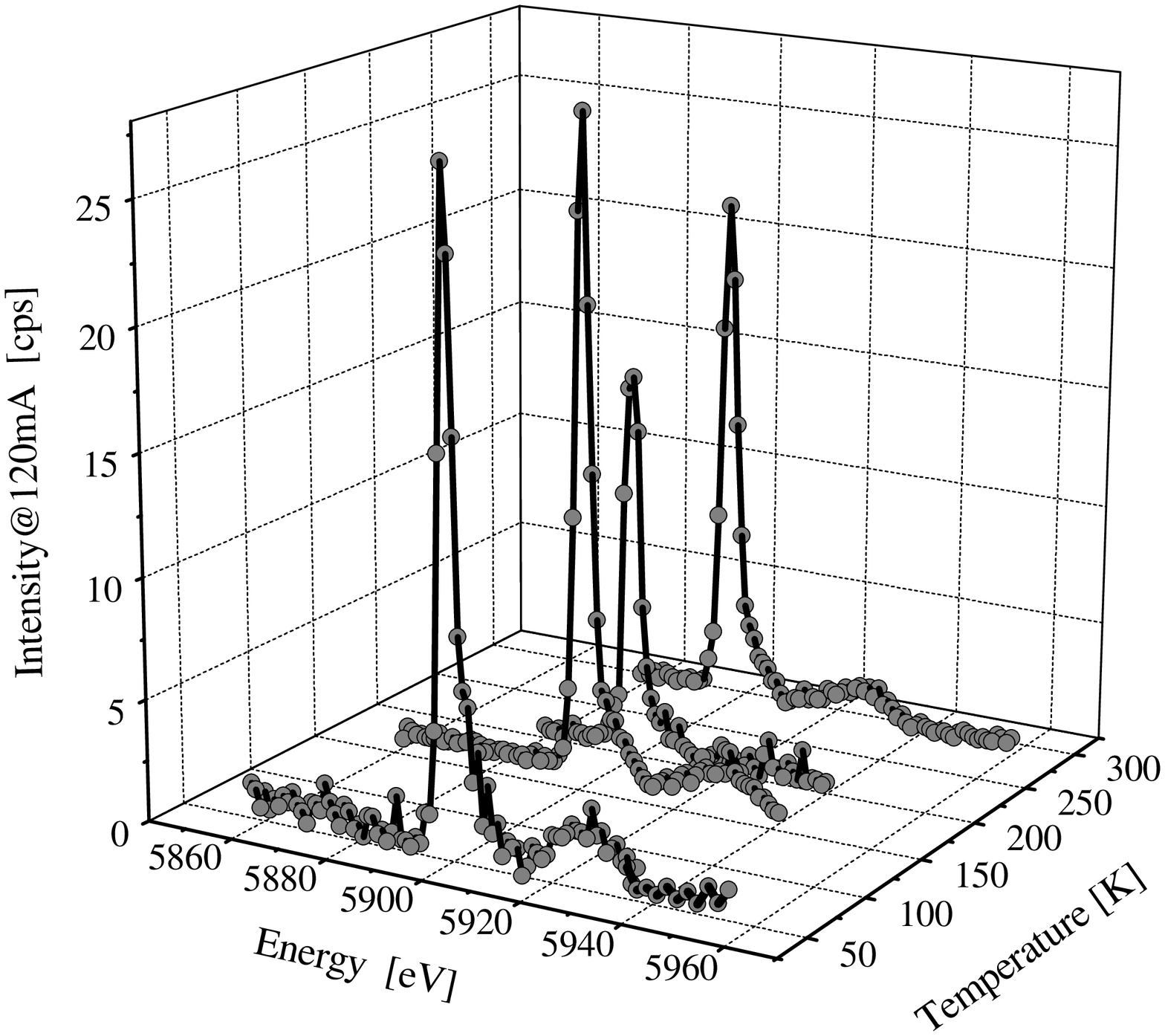}
}}
\caption{Comparison between the energy dependences of the (300)
   reflection taken at 50\,K, 157\,K, 200\,K and 285\,K. The shape of
   the resonance curve is unchanged with increasing temperature whereas
   the intensity is temperature dependent.}
\label{fig:3}
\end{figure}

In Fig.\,\ref{fig:3}, the energy dependences of the (300)-reflection taken at 50\,K, 157\,K, 200\,K and 285\,K are compared. On the one
hand, it can be seen that the shape of the resonance curve is not affected by increasing the temperature; i.e the overall line shape of
the energy dependence is the same in the FMI-phase (T $ <$ \tco $\simeq 155$\,K), the cooperative Jahn-Teller distorted phase (\tco$<$ T
$<$ \tjt$\simeq 280$\,K) and above \tjt.
The data also verify that the resonant scattering is always dominant, while possible non-resonant contributions related to temperature
dependent lattice symmetry changes (see below) can be completely neglected \cite{CoxPRB01}.
On the other hand, it becomes apparent that the intensity of the (300) reflection is strongly temperature dependent as will be discussed
in the following.

\begin{figure*}[t!]
\center{
\resizebox{0.8\columnwidth}{!}{%
   \includegraphics[clip,angle=0]{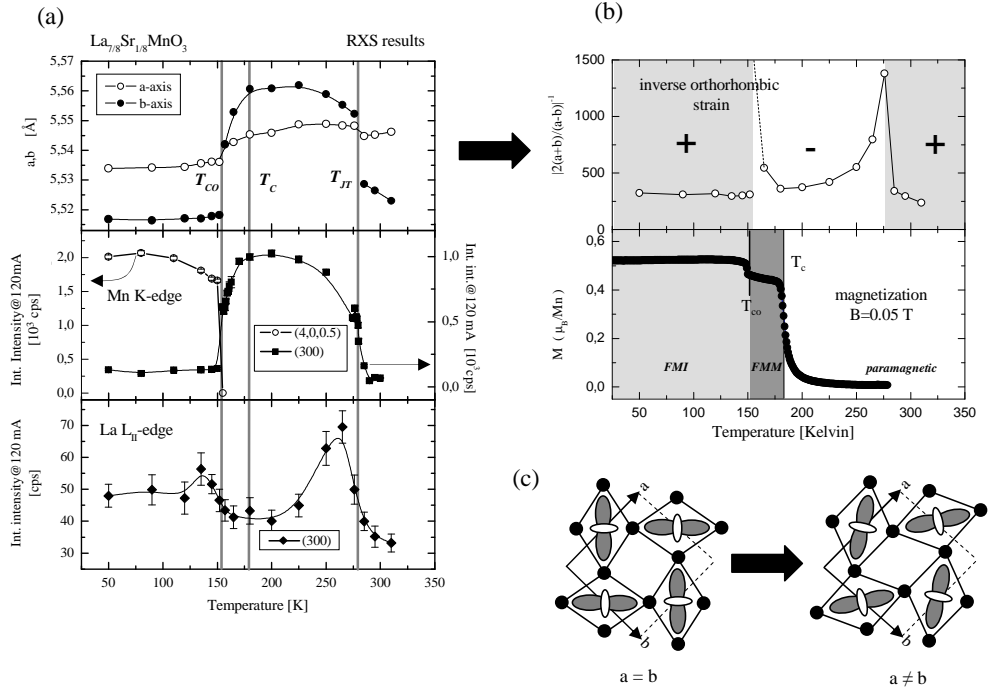}
}} \caption{(a) Comparison of the temperature dependences of the lattice
   parameters ({\it top}), the integrated intensity of the (4,0,0.5) as well as the (300) reflection at the \mnk ({\it middle}),
   and  the integrated intensity of the (300) reflection at the \lii ({\it bottom}). The measurements have been performed with
   increasing temperature. (b) Temperature dependence of $|\gamma|^{-1}$, where $\gamma=2(a-b)/(a+b)$ is the orthorhombic strain
   in the $ab$-plane ({\it top}), and the macroscopic magnetization ({\it bottom}). The +/- signs in the top panel indicate the sign of $\gamma$. (c) Illustration of the connection between the
   $a,b$-splitting, the octahedral distortions, and the octahedral tilts in the cooperative Jahn-Teller distorted phase. Without tilts there is no
   $a,b$-splitting, while a finite tilt around the $c$-axis (perpendicular to the paper plane) results in $a\neq b$. The gray and white lobes represent the occupied $e_g$-states.
    } \label{fig:4}
\end{figure*}
%

\subsubsection{Temperature dependence}

Figure\,\ref{fig:4}\,(a) deals with a comparison between the temperature dependences of the $a,b$-axis, the integrated intensity of the
(4,0,0.5) and the (300) reflection measured at the \mnko , as well as the (300) reflection measured at the \liio . In addition to this,
the temperature dependence of $|\gamma|^{-1}$, where $\gamma=2(a-b)/(a+b)$ is the orthorhombic strain in the $ab$-plane, and the
macroscopic magnetization is given in Fig.\,\ref{fig:4}\,(b).
%
%

%
Upon cooling a cooperative Jahn-Teller distorted and antiferro-orbital ordered phase  develops at \tjt , which is schematically shown in
Fig.\,\ref{fig:4}\,(c) \cite{PinsardJAC97}. The corresponding structural phase transition is clearly reflected by the anomalies of the
lattice parameters $a$ and $b$, which cross at this temperature. At the \mnko , RXS is sensitive to octahedral distortions
\cite{BenedettiPRB01,geckPRB04} and, therefore, the strong increase of the (300) reflection at the \mnk signals the onset of cooperative
octahedral distortions at \tjt, which are connected to an antiferro-orbital ordering. In contrast to the sensitivity to distortions at
the \mnko , the (300) reflection becomes sensitive to changes of the cooperative octahedral tilts for photon energies at the \lii. This
means that the increase of the (300) reflection at the \lii upon cooling reveals changes in the octahedral tilt structure at \tjt.

With further decreasing temperature, the $ab$-splitting and the cooperative Jahn-Teller distortions grow as can be seen in the two upper
panels of Fig.\,\ref{fig:4}\,(a), until both reach a maximum at the ferromagnetic ordering temperature \tc. At the same time, the
concomitant reduction of the (300) intensity at the \lii indicates a coupling between the octahedral tilts and distortions.

The magnetization increase at \tc$\simeq$ 180\,K shown in Fig.\,\ref{fig:4}\,(b), marks the onset of ferromagnetic spin order. In
agreement with the double exchange (DE) mechanism, the onset of ferromagnetic order below \tc \/ is connected to an enhanced charge
carrier mobility, i.e. metal-like behavior \cite{UhlenbruckPRL99}. Therefore, this phase will be referred to as ferromagnetic metallic
phase (FMM) in the following.
As can be observed in  Fig.\,\ref{fig:4}\,(a), the $ab$-splitting as well as the intensity of the (300) reflection at the \mnk is reduced
drastically with increasing ferromagnetic order. The temperature dependence of the (300) at the \mnk implies a suppression of the
cooperative Jahn-Teller distortions in the FMM phase upon cooling.
However unlike the behavior observed around \tjt\/, the suppression of the octahedral distortions is not accompanied by pronounced
changes of the (300) intensity at the \lii below \tc, indicating that the tilt structure hardly changes in the FMM phase.

Finally, the $a$- and $b$-axes cross again, the (300) reflection at the \mnk collapses and new superstructure reflections like the
(4,0,0.5) reflection shown in Fig.\,\ref{fig:4}\,(a) (middle) occur at \tco\/ upon cooling. In a recent publication we showed that the
metal-insulator transition at \tco\/ is connected to a reordering in the orbital sector, leading to the occurrence of new superlattice
reflections \cite{geckPRB04}. More specifically, we revealed that upon cooling the antiferro-orbital order vanishes at \tco\/ and an
orbital polaron lattice is formed \cite{GeckPRL05}. This orbital polaron lattice leads to enhanced ferromagnetic interactions in
agreement with the magnetization jump at \tco \/ that can be observed in Fig.\,\ref{fig:4}\,(b).
The moderate increase of the (300) reflection at the \lii\/ shows that the orbital reordering at the metal-insulator transition is also
connected to changes in the octahedral tilt structure.

At last, we note another interesting experimental observation, which concerns the continuous behavior of the (300) reflection at the \lii
across the first order phase transitions at \tjt \/ and \tco\/. For the transition into the cooperative Jahn-Teller distorted phase a
similar observation has been reported for LaMnO$_3$ \cite{ZimmermannPRB01}. Moreover, a  continuous behavior of the octahedral tilts
across \tjt\/  has also been reported in an neutron diffraction study on LaMnO$_3$ \cite{RudriguezRPRB98}.

\subsection{Discussion}

The data shown in Fig.\,\ref{fig:4} demonstrates the coupling of structural and electronic degrees of freedom  in \LSMO. In the following
the coupling between the octahedral tilts to other degrees of freedom will be discussed in more detail.

First we note, that the $ab$-splitting is determined by two effects in general, namely the octahedral tilts and distortions.
%
%
However,  to leading order the $ab$-splitting can be expressed as
\begin{eqnarray*}
a-b & = & (a-b)_0+ 2^{3/2}\, \delta d\,(1+\mathcal{O}(\alpha_0\,\delta \alpha))\,(\sin \eta_o + \delta \eta ) \\
    &\approx & (a-b)_0+ 2^{3/2}\, \delta d\, \eta_0
\end{eqnarray*}
($(a-b)_0$: $ab$-splitting just above \tjt; $\delta d$: difference between the short and the long Mn-O bond parallel to the $ab$-plane;
$\alpha_0$: tilt angle around $a$- or $b$-axis just above \tjt; $\delta \alpha$: temperature dependent variation of $\alpha_0$; $\eta_0$:
the tilt angle around the $c$-axis above \tjt; $\delta \eta$: temperature dependent variation of $\eta_0$). Assuming moderate variations
of the tilt angles, this means that the temperature dependent changes of $(a-b)$ are mainly given by the distortions of the octahedra.
This conclusion is verified by the experimental data shown in the two upper panels of Fig.\,\ref{fig:4}\,(a), where it can be observed
that the $ab$-splitting is directly related to the octahedral distortions detected by the (300) reflection at the \mnko.

Above \tjt , the octahedra are undistorted on average and the $ab$-splitting can be attributed to the tilting of the octahedra. However,
below \tjt\/ the onset of octahedral distortions ($\delta d \neq0$) changes the $ab$-splitting according to the above equation. The
increasing $\delta d$ leads to a sign change of the orthorhombic strain with increasing $\delta d$ (cf. top panel of
Fig.\,\ref{fig:4}\,(b)) and causes a small $ab$-splitting close to \tjt. With further decreasing temperature, $\delta d$ increases
yielding a large negaive orthorhombic strain. Finally, below \tc\/  the octahedral distortions are reduced and eventually vanish at \tco,
where the sign of the orthorhombic strain changes back to positive.

Focussing on the temperature regime above \tc , it can be observed that the intensity of the (300) reflection at the \lii as a function
of temperature  behaves similar to $|\gamma|^{-1}$, where $\gamma$ is the orthorhombic strain in the $ab$-plane defined above.
In other words, above \tc\ an increased $ab$-splitting is related to a decrease of the (300) intensity at the \liio. This shows that the
variations of the octahedral tilts above \tc\/ are directly coupled to changes of the lattice strain; i.e. the elastic energy of the
lattice. It can therefore be concluded that in the paramagnetic phase the antiferro-orbital ordering and the tilt structure are coupled
mainly via the elastic energy of the lattice.

However, this situation changes below the ferromagnetic transition temperature \tc. Although the $ab$-splitting is strongly reduced in
the FMM phase below \tc\/ ($|\gamma|^{-1}$ diverges), the (300) intensity at the \lii reveals that there are almost no changes of the
octahedral tilt structure. This implies that below \tc, the tilts are not only influenced by changes of the elastic energies. In fact, in
the FM ordered phases the DE mechanism becomes important, which favors larger Mn-O-Mn bond angles, i.e. smaller tilts.

The stabilization of a particular octahedral tilt structure in A$_{1-x}$B$_x$MnO$_3$ is mainly attributed to the coordination of the
A/B-sites \cite{WoodwardActaCryst97a}.  More specifically, different tilt structures lead to different coordinations of these sites and,
hence, yield different energies. However, the experimental data presented here reveals that the octahedral tilt structure in \LSMO\/ is
also strongly affected by the electronic properties related to the correlated Mn:$3d$ and O:$2p$ electrons.
One coupling mechanism between the correlated electron system and the octahedral tilts is given by the transfer integral $t$, which
depends on the Mn-O-Mn bond angles and, therefore, on the octahedral tilts.
%
Since the physics of doped manganites are governed by various competing interactions, the coupling of the electronic degrees of freedom
and the tilt structure observed here, is expected to have a significant impact on
the stability of the various ordered phases.


\subsection{Conclusion}

The energy, azimuthal and temperature dependence of the symmetry forbidden (300) reflection at the \lii has been investigated. At this
energy the (300) reflection is sensitive to changes of the octahedral tilt order. A resonance of the (300) reflection at the \lii has
been observed. Furthermore, the resonant scattering involves a rotation of the beam polarization  (\sigpi-scattering) and displays a
pronounced azimuthal dependence. These characteristic features of the resonant scattering at the \lii\/ are in full agreement with the
model calculation presented in Sec.\,\ref{pheno} and LSDA+U calculations. Therefore, we conclude that the intensity of the (300)
reflection at the \lii\/ reflects changes of the tilt structure.
%
%

The presented experimental results show that the various ordering phenomena observed in \LSMO couple to  the octahedral tilt structure.
%
This conjecture is further supported by the effects of chemical pressure on \LSMO : Upon substituting La by smaller Pr in \LPSMO, the
mean radius of the corresponding lattice site (the so-called A-site) can be reduced in a systematic way, resulting in an increase of the
octahedral tilts \cite{WoodwardActaCryst97a}. Upon increasing the  Pr-concentration $y$ the transition temperature \tco\/ is considerably
reduced, while \tjt \/ strongly increases, manifesting the impact of the octahedral tilting on the ordering phenomena in \LSMO \/
\cite{GeckNJP04}. These results are corroborated by the observed dependences of  \tjt\/ and \tco\/ on applied external pressures
\cite{MoritomoPRB97,ItohPRB97}.

At other doping levels there are as well clear indications for the relevance of octahedral tilts for the physical properties of doped
manganites. For instance, the mean ionic radii of the A-site have also a strong influence on the stability of the charge and orbital
ordered CE-phase in half doped manganites \cite{TokuraScience00}. This points to the important role of the octahedral tilt structure for
the physics of doped manganites in general.

The tilt order is not only of importance in the case of manganites, as already mentioned in the introduction. This order parameter  also
plays a crucial role for the physics of other transition metal oxides like nickelates and cuprates. Also in these cases the RXS technique
is perfectly suited to study octahedral tilt order and its coupling to other degrees of freedom.

\subsubsection*{Acknowledgments}
We would like to thank M. v. Zimmermann for the careful reading of our manuscript and many useful discussions. We are very grateful to H.
Dosch for his support. This work was supported by the Deutsche Forschungsgemeinschaft.

%

%
%

\bibliographystyle{apsrev}



\end{document}